# Rare cosmological events recorded in muscovite mica.

F. M. Russell,
School of Computing and Engineering
University of Huddersfield, HD1 3DH, U.K.

**Abstract.** A study of fossil tracks of charged particles recorded in crystals of muscovite has revealed evidence of rare events of cosmological origin. The events are not compatible with known particle interactions with matter. They were recorded during a period when the crystals were in a metastable state during cooling after growth 13km water equivalent underground. In this state a phase transition can be triggered by low energy events in the range 1eV to 10keV, when the crystals effectively behave as solid-state bubble chambers. At higher energies the chemical etching technique can be used to reveal massive damage to the lattice. The rare events show evidence of interaction with the crystal over a great range of energies. They leave a distinctive record that is easily recognised.

**Introduction**. The search for evidence of exotic events of cosmological origin usually starts with assumptions about possible interactions with ordinary matter. Irrespective of these the detector should offer a large sensitive volume and a moderately long recording time. Ideally, it also should enable detailed study of individual recorded events. An interesting approach looked for fossil evidence of scattering of WIMPs in crystals of muscovite [1,2]. It was based on the possibility that an atom recoiling from a scattering event might cause sufficient damage to a lattice that it could be revealed by the technique of chemical etching. This technique is limited by the extent of damage needed to allow etching and by background recoils generated over geologic time scales from radioactivity, nuclear fission and cosmic radiation. If atomic force microscopy is used to determine the depth of etch pits then the lower limit on recoil energies to give an etchable track is a few tens of keV. This contrasts with the lower limit of about 1eV for recording in muscovite when in the metastable state considered here.

Crystals of muscovite often show visible defects consisting of a hatch-work of black lines lying in the cleavage (001)-plane. Many of these lines lie in principal crystallographic directions at 60º intervals but not all. A study of the properties of these exceptions showed that some were the fossil tracks of charged leptons. In particular, some were the tracks of positrons emitted from the isotope $^{40}$K that occurs in the monatomic sheets of potassium forming part of the crystal structure. It was found that the recoil of the nucleus arising from the dominant beta decay channel created a mobile lattice excitation called a quodon. These quodons can trap a charge and propagate unimpeded along chains of potassium atoms for great distances. They move at ~3km/s and are the cause of the majority of lines lying in the 60º directions. Evidence also was found for fossil tracks due to e-p showers [3]. These showed that the tracks were recorded after the crystal had grown but the temperature was still above 700K, which allowed migration of atoms to form the black lines. The recording process operating in the metastable state does not depend on ionisation. It arises from a phase transition triggered by the presence of a positive charge when the crystal is in a metastable state during cooling. In this state the lattice needs nucleation sites to expel excess iron to form the black ribbons of magnetite. The sensitivity of this process is shown by the lower limit of energy of a quodon of about 1eV for it to be recorded. In effect, the crystals behave as a solid-state bubble chamber.

In common with many silicate minerals the composition of muscovite is variable, as is the impurity content. This results in variability of the recording sensitivity for different causes of the nucleation sites. One consequence is the extent to which the initial delineation of a track at nanometre-scale subsequently grows by lateral accretion to become several orders of magnitude wider. This decoration does not change the occurrence, origin, orientation and physical properties of the initial fossil track but affects their visibility. Clearly, in a perfect crystal the recording process cannot operate. The most finely decorated tracks are caused by charged relativistic particles, such as muons, passing through crystals with little iron and few other impurities. The tracks of quodons are seldom found in these crystals even though they are created copiously in beta decay of $^{40}$K. As the iron content increases the recording sensitivity also increases allowing quodons to be recorded as well. Eventually, when the iron content has risen to about 6 atomic percent the decoration is extensive. It is then possible for laterally unstable kink-like excitations of the lattice, created in atomic cascades caused by nuclear scattering events, also to be recorded. Their fossil tracks are distinctive because of their fan-shape. Almost all the decoration is composed of the black mineral magnetite in the form of very thin ribbons. These ribbons are intrusive in the potassium (001)-planes, where the lattice is weakest and easily cleaved.

Figure 1 shows the scan of a sheet of muscovite containing a nuclear star and illustrates the detail that can be recorded by the phase transition recording process. The sensitivity is similar to that of photography.

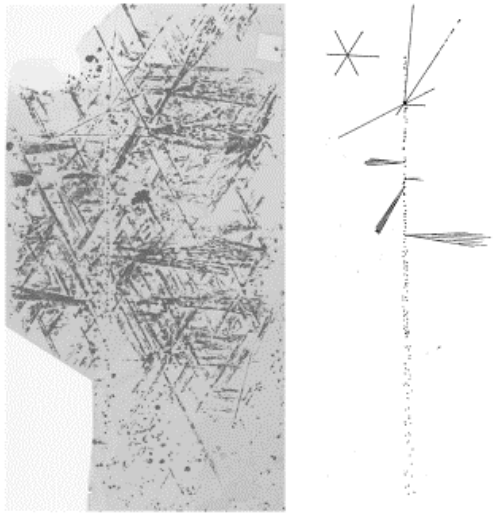

Figure 1. Scan of sheet of muscovite showing the fossil tracks of charged particles.

The diagram identifies the relevant parts of the fossil tracks resulting from a nuclear star. The directions of the principal atomic chains are shown. Most of the tracks lying in these directions are due to quodons. Some tracks can be channelling relativistic particles but these usually show fans from nuclear scattering events, which quodons cannot create. The direction of flight of the particle causing the star is unknown. It results in at least eight tracks. When tracks lie in the (001)-plane they are usually continuous. For those moving at an angle to this plane they will intersect the potassium sheets, where the recording occurs, at separated points. The long vertical chains of dots are of this type. The fan shaped patterns are caused by nuclear scattering events that produced atomic cascades in which kink-like lattice excitations are created. These fans are clustered around the principal crystal directions. The sheet of mica is 15cm x 29cm.

The first rare event was found in 1971. By 1993 evidence had been found suggesting the existence of a mobile lattice excitation created in beta decay of $^{40}$K. called a quodon. Numerical modelling indicated that atoms might be ejected by inelastic scattering of quodons at a remote point after travelling >$10^7$ atoms from the point of creation by irradiation with alpha particles. This was observed and published in 2007 [4]. This finding led to the prediction in 2016 of infinite charge mobility in muscovite of charge carried by quodons. This was verified by experiment and published in 2017 [5]. It provided independent support for the existence of quodons and the sensitive recording process in muscovite. By then a second rare event had been found in a total of ~0.03m$^3$ or 75kg of muscovite. After measurements on, and identification of the cause, of nearly a million fossil tracks the rare events remain unexplained.

**Description of rare event**. The event is shown in figure 2 as a scan of the muscovite sheet and then after cleaving into four thinner sheets. The diagram illustrates the main features in each sheet. It is reasonable to assume that the event originated at the central dot from which two oppositely directed sprays of dots emerge plus a long jet-like array of dots. The central massive dot and the emerging sprays and jet are not restricted to one K-plane but extend into the bulk material over >3.8x$10^5$ K-planes. The central dot and the smaller ones are composed of magnetite. None of the recorded tracks or curvilinear arrays of dots lie exactly in principal crystallographic directions. The second picture shows the central region at larger scale: the area of the sheet covered by the picture is 4 x 2.7cm$^2$.

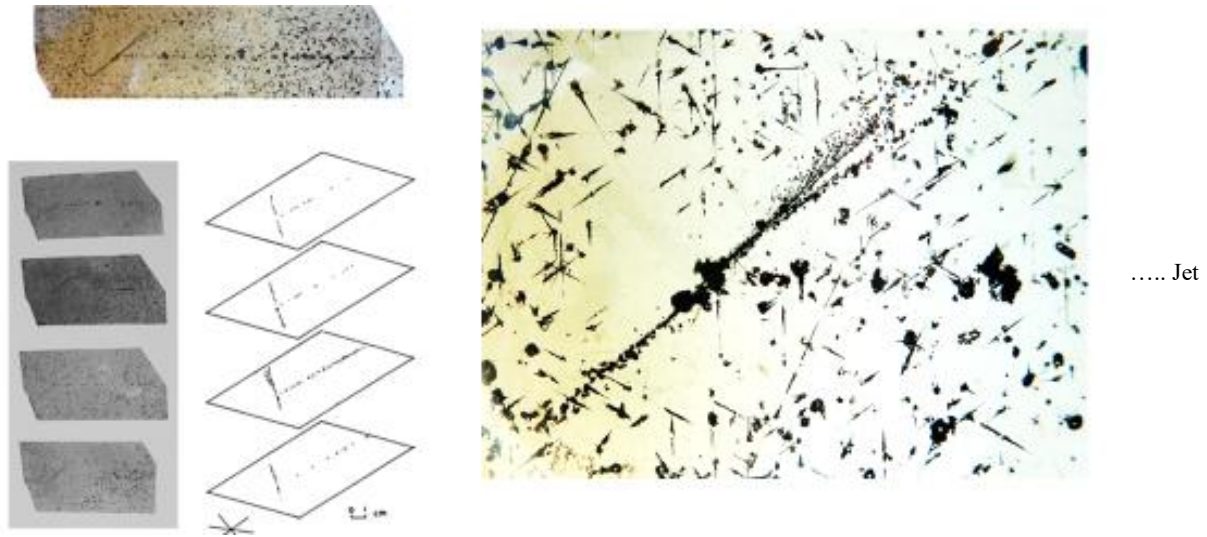

..... Jet

Figure 2. Pictures showing the rare event. The event is distributed in the z-direction normal to the (001)-plane of the sheets. There are less than average number of dots and short tracks in the region surrounding the event. The absence of any fan-shaped tracks in any of the layers puts limits on the cause of the event as fans are copiously produced by relativistic heavy ions. The length of the top mica sheet is 14cm.

The decoration of the centre region with magnetite is continuous in the z-direction. It contains of order $10^{18}$ atoms and caused a major restructuring of the lattice. If the nett energy needed to restructure the atoms in the core is only 0.01eV per atom then the input energy to the crystal to create the central dot alone is of order $10^{16}$ eV or $10^4$ TeV. Since the cause of the event is unknown there could be other kinds of interaction, such as annihilation, creating a core void that is then filled with magnetite. Although the recording and decoration processes are exothermic the majority of the energy released by these processes is available only after the cause of the event has ended its interaction with the lattice. Muscovite contains some heavy elements that on fission give by-products of micron range. They would certainly create atomic cascades but no fans are associated with the rare events. The roughly linear alignment of dots in the long jet is contained within an opening angle in the (001)-plane of about 1°. A puzzling feature is that the long jet appears to have a distributed origin at different depths in the core instead of at one point. The intensity of decoration on the two smaller jets or sprays is not equal. The absence of any fans and of any continuous tracks of charged particles emanating from the core is incompatible with a nuclear star origin. None of the jets lie exactly in principal crystallographic directions and so are unlikely to be influenced by, or be an artefact of, the lattice. The structure close to the core has several puzzling features: there appears to be a lateral offset in the start of the two opposing sprays. The existence of the long jet, the opposing sprays of dots and the central magnetite mass are incompatible with an included crystal, for example, of garnet or zircon.

**Discussion.** The isolation of the event from the surface of the crystal, the absence of fractures leading to a crystal edge and the localised disruption of the recording process for other events all point to a massive release of energy in a crystal buried 13km water equivalent underground. This extent of shielding points to a cosmological origin. Extremely high energy neutrinos are a source of high energy muons. The stopping power for muons in mica can reach of order 100MeV/cm, depositing about 10MeV in the 0.1cm thickness of the mica sheet [6]. This is far too small to have caused the observed damage. The existence of the central magnetite core penetrating through thousands of layers has not been seen elsewhere. It raises the matter of where the atoms that originally occupied the volume of the central core went. If the event was the only example then a fortuitous confluence of causes might be constructed but the existence of a second similar event makes this unlikely. Part of a possible third event also has been found. Would such events be seen in other detectors? The rate of occurrence of the observed events, two in 0.03m$^3$ in 500 years, is sufficiently low that it could be missed. They consist mainly of intense structural damage and a long-range triggering of the recording process in a cone of narrow opening angle. Structural damage would not be observable in most crystals and hidden in metals. In liquids the structural damage would not persist. The rare events were found in relatively thin sheets of the crystal that were cleaved from larger crystals and then randomly mixed at the mine. It would be informative to study the extent of the event in the z-direction.

**Conclusion.** A detailed study of 0.03m$^3$ of muscovite crystals containing over a million fossil tracks of charged particles and mobile lattice excitations called quodons yielded two events that showed exotic features incompatible with known high energy particles. The events show a core region containing about $10^{18}$ atoms where the normal lattice has been replaced by magnetite. Two diagonally opposed narrow sprays of dots of magnetite originate from the core, indicative of local ionisation. These sprays extend up to 3cm from the core. They are not aligned with the prime crystallographic directions. In addition, there is a long-range jet-like spray of magnetite-decorated ionisation dots extending from the core for 10cm with an opening angle of about 1º. This jet-like spray does not originate from a single point in the core but from different points in the $4 \times 10^4$ sheets of unit-cell thickness. In one event the damage to the crystal in the core region caused local fracturing of the crystal, which is localised and does reach to a crystal edge. The region of the crystal surrounding each event contains the usual background of fossil tracks, showing that the properties of the crystal during the recording stage was normal. The events are inconsistent with an origin due to nuclear stars, heavy ions, e-p showers and fission of heavy nuclei. The replacement of the crystal in the core region with magnetite implies a residual distributed input of positive charge. This cannot be due to passage of a swift heavy ion because there are no fan-shaped fossil tracks due to lattice-kinks that are created in atomic cascades.

**Acknowledgement.** It is a pleasure to acknowledge helpful discussions with J C Eilbeck and P A Lindsell.